\documentclass[
reprint,
groupedaddress,
showpacs,
showkeys,
amsmath, amssymb,
aps,
prx,
]{revtex4-2}

\usepackage{xcolor}
\usepackage{nicefrac, xfrac}
\usepackage{tabularx}
\usepackage{graphicx}
\usepackage{dcolumn}
\usepackage{bm}
\usepackage[colorlinks=true, allcolors=blue]{hyperref}
\usepackage{comment}
\usepackage[normalem]{ulem}
\graphicspath{{figures/}}
\usepackage[separate-uncertainty=false]{siunitx}
\usepackage{braket}
\usepackage{xspace}
\usepackage{xstring}

\usepackage{adjustbox}
\usepackage{multirow}

\newcommand{\na}{$^{23}$Na\xspace}

\newcommand{\red}[1]{\textcolor{black}{#1}}

\newcommand{\blue}[1]{\textcolor{blue}{}}
\newcommand{\suppress}[1]{\textcolor{brown}{}}

\newcommand*{\aref}[1]{%
	\IfBeginWith{#1}{eq:}{Eq.\,\eqref{#1}}{}
	\IfBeginWith{#1}{fig:}{Fig.\,\ref{#1}}{}%
	\IfBeginWith{#1}{tab:}{Table~\ref{#1}}{}%
	\IfBeginWith{#1}{appendix:}{Appendix~\ref{#1}}{}%
	\IfBeginWith{#1}{sec:}{Section~\ref{#1}}{}%
	}

\begin{document}

\title{Progress toward a zero-magnetic-field environment for ultracold-atom experiments}

\date{\today}

\author{C. Rogora$^1$}
\author{R. Cominotti$^1$}
\author{C. Baroni$^{1,2}$}
\author{D. Andreoni$^1$}
\author{G. Lamporesi$^{1,3}$}
\email{giacomo.lamporesi@ino.cnr.it}
\author{A. Zenesini$^{1,3}$}
\author{G. Ferrari$^{1,3}$}

\affiliation{$^1$Pitaevskii BEC Center, CNR-INO and Dipartimento di Fisica, Universit\`a di Trento, 38123 Trento, Italy}

\affiliation{$2$Institute for Quantum Optics and Quantum Information (IQOQI), Austrian Academy of Sciences, 6020 Innsbruck, Austria}

\affiliation{$^3$Trento Institute for Fundamental Physics and Applications, INFN, 38123 Trento, Italy}

\begin{abstract}

The minimization of the magnetic field plays a crucial role in ultracold gas research. For instance, the contact interaction dominates all the other energy scales in the zero magnetic field limit, giving rise to novel quantum phases of matter. However, lowering magnetic fields well below the mG level is often challenging in ultracold gas experiments.
In this article, we apply Landau-Zener spectroscopy to characterize and reduce the magnetic field on an ultracold gas of sodium atoms to a few tens of $\mu$G. The lowest magnetic field achieved here opens \red{the way} to observing novel phases of matter with ultracold spinor Bose gases.

\end{abstract}
\maketitle

\section{INTRODUCTION}

Controlling magnetic fields is critical in many contexts involving fundamental and applied physics experiments and quantum technologies. 
Often, the performance of a measurement critically depends on the stability against fluctuations of the background magnetic field, as it happens in electron microscopy experiments \cite{Mulvey1962,Ruska1987,Krivanek2008}, atom interferometry \cite{Varenna2019}, nuclear magnetic resonance \cite{Mansfield1987}, atomic clock experiments \cite{Clairon1991, Gibble1993, Ludlow2015}, and ultracold gases experiments involving coherently-coupled condensate mixtures \cite{Nicklas2015,Cominotti2024}.

In other contexts, such as zero-field nuclear magnetic resonance \cite{Ledbetter2008, Tayler2017}, the constraint relates to the magnitude of the magnetic field, which generally should be minimized and, depending on the type of measurement, needs to be kept below the threshold values. Historically, the measurement of relatively small magnetic fields is achieved using SQUIDs in a cryogenic environment \cite{Romani1982}, exploiting atomic coherence in room temperature gas cells \cite{Dupont-Roc1969, Budker2002, Budker2007}, or atomic spin-alignment \cite{Meraki2023}, with applications in diverse fields, including biomedical imaging \cite{Brookes2022, Sandler2012}, metal detection \cite{Bevington2019, Rushton2022}, and material characterization \cite{Romalis2011}. Nowadays, a wide variety of experimental platforms and techniques are available, as described, for instance, in Ref.\,\cite{Mitchell2020}. In the context of magnetometry, cold gases have been consistently employed as magnetic field sensors \cite{Isayama1999, Cohen2019}, in various cases with micrometric scale spatial resolution \cite{Higbie2005, Wildermuth2006, Vengalattore2007, Eliasson2019, Yang2020}\red{\, exploiting the enhanced sensitivity of spinor condensates to magnetic field inhomogeneities \cite{Witkowski2013}}.

Finding a way to reduce the magnetic field, control it with high accuracy, and be able to measure such small values would pave the way for the investigation of new physical phenomena, such as the zero-magnetic-field physics of spinor condensates, {\it i.e.}, condensates with a vector order parameter. Spinor condensates can develop different configurations, and their ground state depends on the strength of the interactions between the internal states and on the strength of the externally applied magnetic field, which removes the degeneracy between different spin states \cite{StamperKurn2013}. 
In the majority of the experiments, the magnetic field is typically large enough that the corresponding magnetic energy splitting between spin states dominates over the spin-dependent interaction energy. 
In such an experimental configuration, spinorial systems with a nonzero and fixed magnetization can be investigated \cite{Chomaz2023}. The complementary case where the magnetization is set to zero and evolves in the absence of an external magnetic field is also very interesting and yet unexplored, and this work aims at creating the conditions to investigate it experimentally. 

A regime where the two energies are comparable has been studied with dipolar chromium gases in the presence of magnetic fields of a few tenths of a mG, thanks to the large magnetic moment of chromium atoms \cite{Pasquiou2011}.

In ultracold gases of alkali atoms, the contribution of the long-range magnetic dipole interaction is typically small \cite{Fattori2008}, which makes the spin-dependent interaction dominated by the spin-dependent contact interactions. This further reduces the threshold magnetic field below which interesting and unobserved spinor phases are expected with respect to the chromium case \cite{StamperKurn2013,JimenezGarcia2019, Hamley2012, Zibold2016, Evrard2021}. Reducing the amplitude of the magnetic field may allow, for instance, for the observation of fragmentation in sodium condensates without manipulating the energy levels to establish the degeneracy among the Zeeman sublevels \cite{Evrard2021}. Besides, it is also relevant in the context of dipolar magnetic gases where the magnetic interaction becomes crucial and interesting phenomena can develop such as spontaneous circulation \cite{Kawaguchi2006}.

The strength of the magnetic field below which these novel phases appear is known, in the case of the zero spatial mode approximation \cite{Hamley2012, Zibold2016, Evrard2021}, to be on the order of a few hundreds of $\mu$G. However, this threshold is expected at even lower magnetic fields in the case of spatially extended condensates. Here, the experimental challenge relies on the difficulty of controlling the magnetic field stability at the $\mu$G level and reaching and maintaining such small magnetic field values over the whole extension of the atomic sample both during a single experimental sequence and within different runs. 

This work presents an experimental technique for minimizing the magnetic field at the 10-$\mu$G level.
Since it is \red{technically difficult} to integrate an external device in the vacuum chamber of ultracold atom experiments to characterize and minimize the field, developing an independent technique for the magnetic field characterization using the trapped atoms as sensors becomes necessary. In particular, we describe an application of Landau-Zener (LZ) spectroscopy \cite{Band2019} over an atomic gas of sodium. Reaching such a low field is possible thanks to the presence of a magnetic shield \cite{Farolfi2019}, which demonstrated its efficiency in stabilizing the field in several previous works \cite{Farolfi2021QT,Farolfi2021,Cominotti2023,Zenesini2024}.
The results presented in the following open the way to studying the zero-magnetic-field ground state of an $F$=1 system.    

In Sec.~\ref{SectionII}, we describe the experimental platform. Section~\ref{SectionIII} contains the theoretical framework and experimental protocols. In Sec.~\ref{SectionIV} we discuss the results, while in Sec.~\ref{SectionV} we report concluding remarks and outlooks. 

\section{Experimental platform}
\label{SectionII}

The experimental platform consists of a bosonic gas of \na atoms, trapped in the elongated optical trap potential generated by a single far-detuned infrared (1064\,nm) laser beam, and therefore equal for all spin components. A thermal sample is loaded in the optical trap and presents a Gaussian spatial density distribution in all three directions. In the optical trap ($\omega_\mathrm{x}/2\pi \sim 10$~Hz, $\omega_\mathrm{y,z}/2\pi \sim 1$~kHz,), the cloud has an elongated shape with a 1:100 aspect ratio, having the long axis along $x$ and the short axes along $y$ and $z$. 

\begin{figure}[b!]
 \centering
  \includegraphics[width = 1.0\columnwidth]{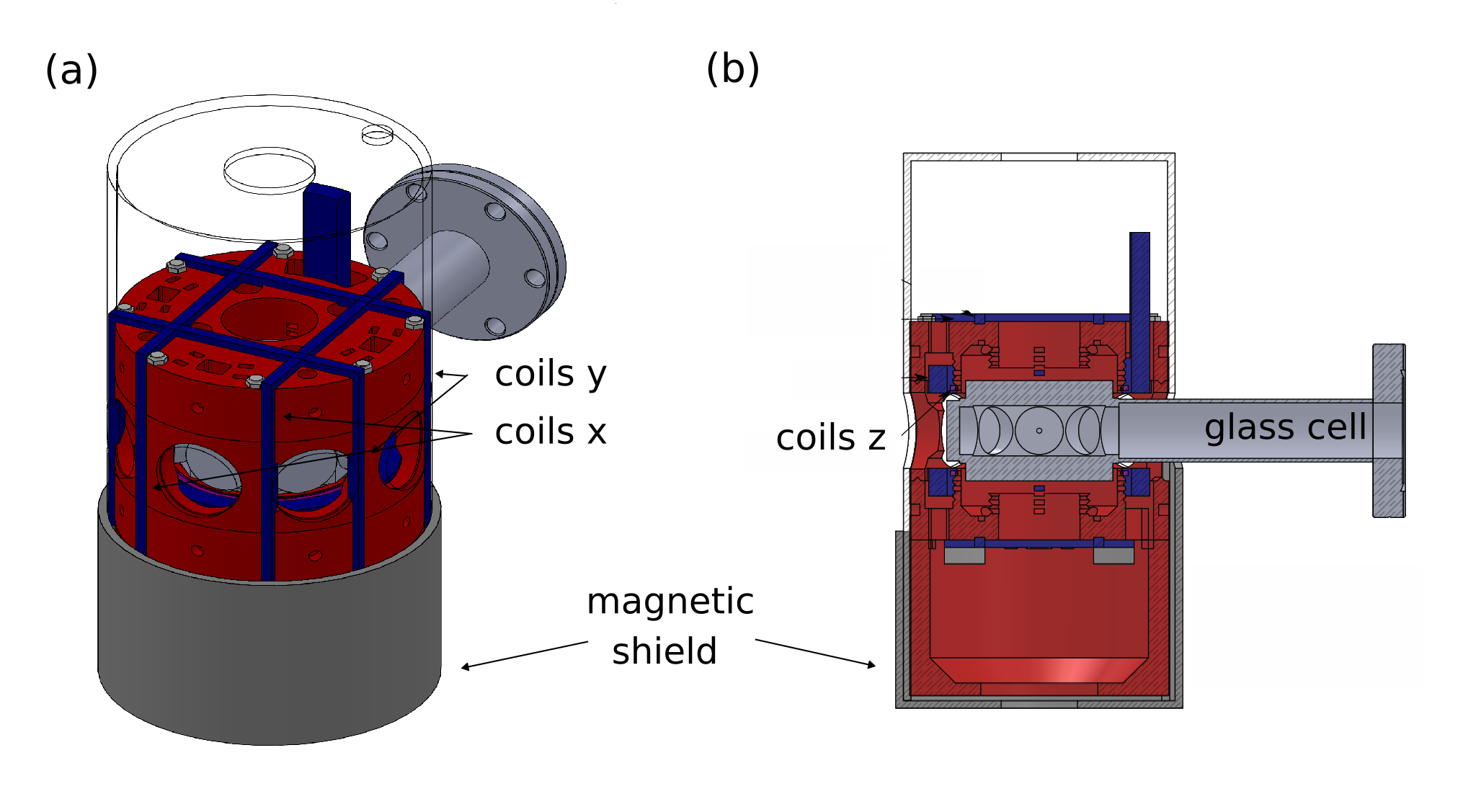}
  \caption{Schematics of the core setup. (a) The innermost layer of the magnetic shield (dark gray and transparent) contains a 3D-printed plastic support structure for the coils (red), the coils themselves (blue), and the glass cell (light gray). (b) $yz$ section of the apparatus.}
   \label{fig:fig1}
\end{figure}

The sample is prepared inside a magnetic shield, which guarantees the stability of the magnetic field at the $\SI{}{\mu G}$ level \cite{Colzi2018,Farolfi2019}. Pairs of magnetic coils inside the innermost layer of the magnetic shield allow for the application of controllable magnetic fields in each of the three spatial directions. Figure~\ref{fig:fig1} schematically represents the apparatus and shows the glass cell, coils, and magnetic shield. 
The applied magnetic field's stability is ensured using high-stability current supplies [Stanford research system (SRS) LDC501 for the longitudinal field and Delta Elektronica ES 015-10 with 10:1 current dividers for transverse fields]. 
Ramping from positive to negative values of the longitudinal field within the same experimental run is achieved using two unipolar current supplies with opposite orientations arranged in parallel. A master SRS is set to the steady drive of 100 mA ($\sim$ 245 mG), while a second SRS introduces a tunable current in the opposite direction. 

\section{Measurement scheme}
\label{SectionIII}
Spectroscopic methods are typical solutions to measure a magnetic field. They consist of interrogating an atomic two-level system with constant radiation at different frequencies for a given time and recording the resulting energy spectrum. In such measurements, the Fourier broadening does not represent a limitation at high magnetic fields when the Zeeman splitting ($\sim 700$ kHz/G for atomic \na) is much larger than the resonance linewidth. Conversely, when approaching the $\mu$G regime, techniques such as radio-frequency spectroscopy are difficult to apply, given the unresolved Zeeman structure, and the difficulty of defining the polarization of the radio-frequency or microwave coupling fields.

An alternative method to characterize the magnetic field around the null value relies on ramps of the magnetic field amplitude and directions, taking advantage of the adiabatic/diabatic dynamics of the atomic spin rotation. At low magnetic field, {\it i.e.}, when the Larmor frequency is of the order or lower than the velocity of rotation of the magnetic field direction, LZ theory \cite{Landau1932,Zener1932,Kofman2023} applies, resulting in a powerful tool for the magnetic field characterization.

We developed two protocols based on LZ sweeps on the $z$ (taken as a quantization axis) component of the field. 
The first one aims to minimize the magnetic field in the transverse $xy$ plane and is implemented by ramping (or sweeping) the $z$ field component from positive to negative finite values. The second protocol involves a ramp with a variable endpoint to find the minimum field along $z$. 

In the following, we will use the magnetic field nomenclature presented in Tab.\,\ref{tab:tab1}. 

\begin{table}[h]
    \centering
    \begin{tabular}{|l |l|}
    \hline
         $B$ & Magnetic field modulus \\

         $B_i$ ($i=x,y,z$) & Actual magnetic field components \\

         $B_\perp$ & Transverse field \\

         $B_z$ & Longitudinal field \\

         $B_{i,\mathrm{coils}}$ & Field induced by the $i$ coil\\

         $B_{i,\mathrm{ramp}}$ & Starting  $B_{i,\mathrm{coils}}$ of the LZ ramps\\

         $B_{i,0}$ & Optimal $B_{i,\mathrm{coils}}$ from fit \\

         $B_{z,c}$& Field set during condensation \\

         $B_\mathrm{z,fin}$& Final value of longitudinal field protocol \\
    \hline
    \end{tabular}
    \caption{Magnetic field nomenclature.}
    \label{tab:tab1}
\end{table}

\subsection{Energy levels and Landau-Zener theory}

\begin{figure}[t!]
 \centering
  \includegraphics[width = 1\columnwidth]{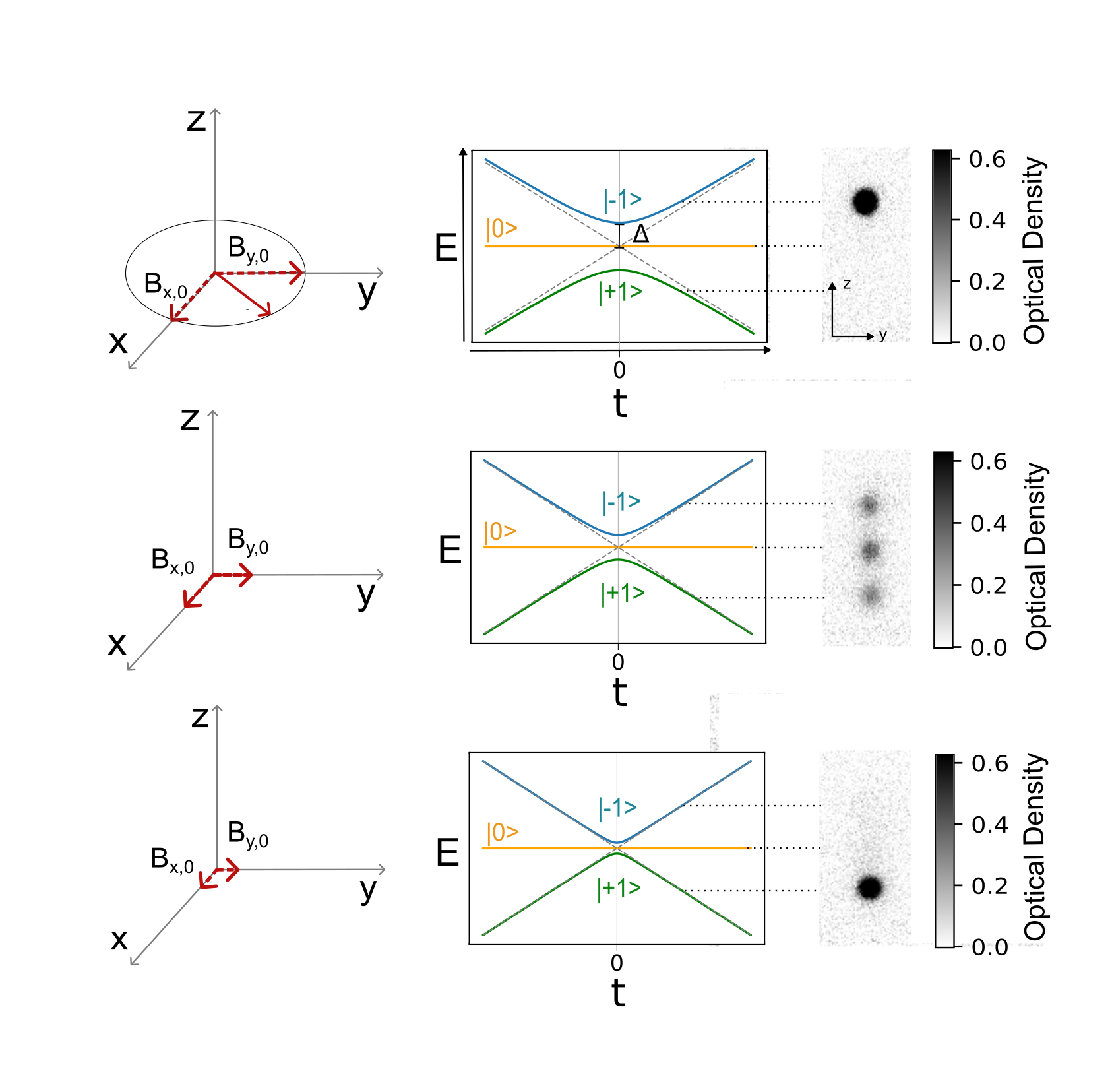}
  \caption{Transverse field and LZ ramps. For each row, the left panels show a schematic view of the transverse magnetic field components ($B_\perp$ decreases from top to bottom). Central panels show the energy of the three states $m_F = 0,\pm 1$ as a function of time for the corresponding $B_\perp$. The right panels show the relative populations of the atoms in the three states at the end of the ramp, obtained experimentally and measured through Stern-Gerlach imaging. For the three different values of $B_\perp$ we obtain full adiabatic transfer (upper panel, large $B_\perp$), partial transfer (central panel), and complete diabatic transfer (lower panel, low $B_\perp$).}
   \label{fig:fig2}
\end{figure}

The hyperfine ground state of sodium has a total angular momentum $F=1$ with three magnetic sublevels $m_F=0,\pm 1$ defined as the projection of $F$ along the $z$-axis. The energy of the three states can be estimated with high accuracy using the Breit-Rabi formula \cite{Breit1931} once the magnetic field value is known, and, in the small field limit, a linear dependence on the field  (first order Zeeman regime) is expected.
Let us consider $z$ as the quantization axis and the magnetic field with a finite transverse component $B_\perp=(B_\mathrm{x}^2+B_\mathrm{y}^2)^{1/2}$.
If $B_\mathrm{z}$ is linearly ramped with a slope $\dot{B_\mathrm{z}}$, from large positive to large negative values passing across zero, the energy of the states as a function of time is the one shown in Fig.\,\ref{fig:fig2}. At $B_\mathrm{z}=0$, the magnetic field is equal to $B_\perp$, which introduces an avoided crossing between the states.

The time-dependent Hamiltonian of the system for the wavefunction $\{\psi_{-1},\psi_0,\psi_{+1}\}$ reads, \red{as in \cite{Carroll1986, Band2019}},
\begin{equation}
H(t) = g_F \mu_B
\begin{pmatrix}
\alpha t&\Delta&0\\
\Delta&0&\Delta\\
0&\Delta&-\alpha t\\
\end{pmatrix},
\end{equation}
where $g_F=-1/2$ is the Land\'e factor, $\mu_B$ is the Bohr magneton, and $\Delta=B_\perp /\sqrt{2}$ acts as a coupling between the states and defines the gap in the energy levels shown in \aref{fig:fig1}. Here $t=0$ is defined as the instant when
$B_z=0$. The parameter $\alpha$ originates from the time derivative of the Larmor frequency at the energy levels crossing 

\begin{figure}[b!]
 \centering
  \includegraphics[width = 1\columnwidth]{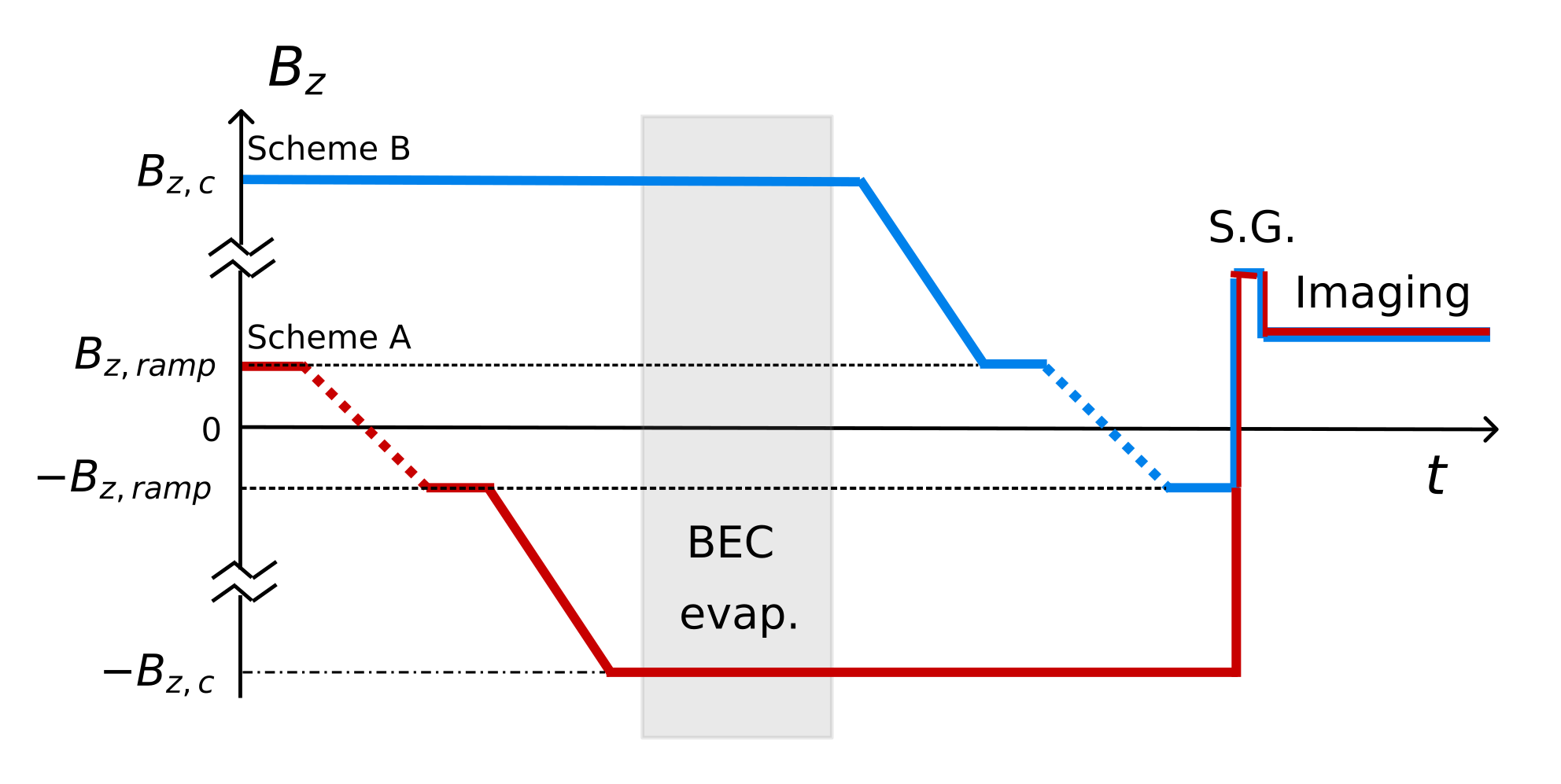}
  \caption{Transverse field minimization protocol. $B_\mathrm{z}$ is shown as a function of time. Scheme A (red line) consists in a ramp from $B_\mathrm{z,ramp}$ to -$B_\mathrm{z,ramp}$ performed on a thermal cloud, then $B_\mathrm{z}$ is brought down to $-B_\mathrm{z,c}$ in 50~ms before the evaporation in the optical trap.  Scheme B (blue line) starts with the condensation at a finite value of the field $B_\mathrm{z,c}$, then after the evaporation, the field is ramped down to $B_\mathrm{z,ramp}$ in 50~ms. Here the condensate experiences the field ramp across zero, from $B_\mathrm{z,ramp}$ to $-B_\mathrm{z,ramp}$. In both cases, the dotted line corresponds to the field crossing the minimum value with a tunable ramp duration. The shaded gray area denotes the time during which the sample is evaporated to obtain a BEC.}
   \label{fig:fig3}
\end{figure}

\begin{equation}
\alpha = \left| |B(t)| \frac{\partial}{\partial t} \left( \arctan \frac{\dot{B}_\mathrm{z} t}{|B(t)|} \right) \right| _{t=0}= |\dot{B}_\mathrm{z}|\red{,}
\end{equation}

\red{which relates the $B_z$ ramps with the variation of the Larmor precession of the atoms around the applied magnetic field.} 

\begin{figure*}[t!]
 \centering
  \includegraphics[width = 0.8\textwidth]{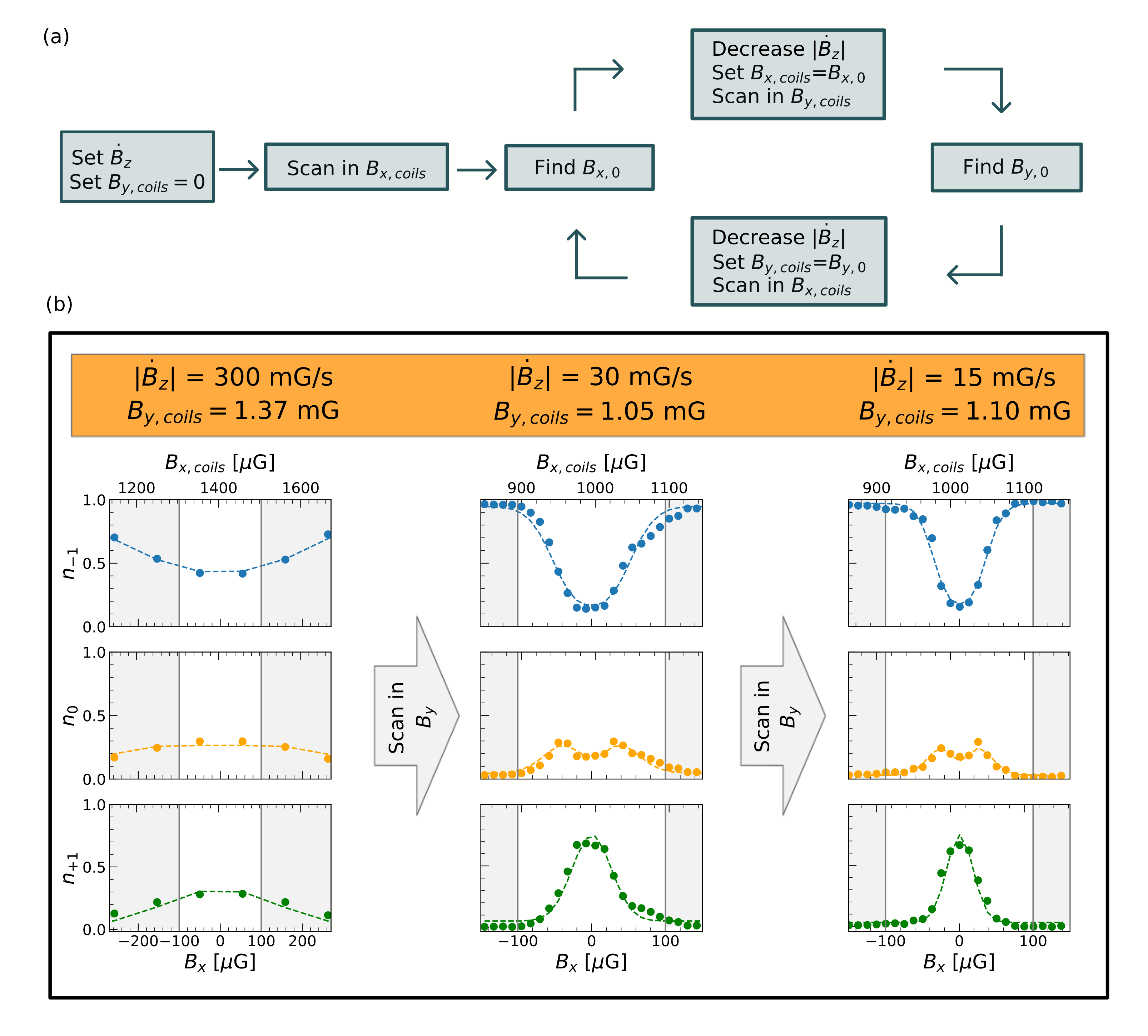}
  \caption{Transverse field determination. (a) Schematic representation of the iterative measurement process; see main text. (b) Relative populations of the three Zeeman levels $m_F=-1, 0, +1$ (from top to bottom) for decreasing ramp velocities $\red{|\dot{B}_z|}$ (from left to right). All panels show the experimental data (dots) and the fitted function (solid line) plotted as a function of the transverse field in the $x$ directions. For each panel, the upper axis shows the field's value generated by the coils, while the lower axis is shifted according to the fit result. The speed of the ramp is specified in the orange box as well as the value of the compensation field in the $y$ direction used for each scan. Note the offset of the compensating field in the first column, whose data were acquired with a few weeks delay with respect to the other two data sets.}
   \label{fig:fig4}
\end{figure*}

While the atoms are initially prepared in the $m_F=-1$ state at large (as compared to $B_\perp$) and positive values of $B_\mathrm{z}$, ramping $B_\mathrm{z}$ in time may induce the transfer to different $m_F$ states according to the analytical model discussed in \cite{Band2019}. In the following, we apply the results of \cite{Carroll1986} as we measure the $m_F$ population distribution at long times after the inversion of $B_\mathrm{z}$:

\begin{align}
P_1&= p^2,\label{eq:P1}\\
P_0 &= 2p(1-p),\label{eq:P0}\\
P_{-1} &= (1-p)^2,
\label{eq:Pm1}
\end{align}
with 
\begin{equation}
p = e^{-2\pi |g_F| \mu_B\frac{\Delta^2}{\red{2}\hbar \red{|\dot{B_z}|}}}
\label{eq:p}
\end{equation}
being the LZ transfer probability. From Eq.~(\ref{eq:P1}) and Eq.~(\ref{eq:p}), one finds that a complete diabatic transfer of the population from the $m_F = -1$ to the $m_F = +1$ state takes place when $p\approx1$ (large ramp speed $\red{|\dot{B}_z|}$ or small gap $\Delta$), while adiabaticity is preserved for small $\red{|\dot{B}_z|}$ and large $\Delta$. This results from implementing the adiabatic condition to the spin dynamics in a rotating magnetic field. In other words, adiabaticity is fulfilled when the variation of the magnetic field direction is smaller than the Larmor frequency, {\it i.e.}, ${|\dot{B_\mathrm{z}}}/{B}|\ll |{g_F\mu_B B}/{\hbar}|$. For instance, the probability of the transition to $m_F=1$ has a Gaussian distribution with RMS width 
\begin{equation}
\sigma_B=\sqrt{\frac{\hbar \dot{B}_\mathrm{z}}{2 \pi |g_F| \mu_B }},
\end{equation}
which is the width used in the following.

\subsection{Experimental protocols}

\subsubsection{Transverse field minimization}

The goal is to find the optimal current values for each coil to compensate for the residual transverse magnetic field (which is not screened by the magnetic shield or due to the shield's permanent magnetization).

The protocol to minimize the transverse field amplitude $B_\perp$ starts with a thermal atomic sample by setting the values of the currents in the coils for the transverse directions, generating the fields $B_\mathrm{x,coils}$ and $B_\mathrm{y,coils}$. 
In the following, we discuss the two experimental schemes depicted in \aref{fig:fig3}.

Scheme A (red line) consists in changing $B_\mathrm{z}$ from $B_\mathrm{z,ramp}$ to $-B_\mathrm{z,ramp}$ with a linear ramp of variable duration $\Delta t^*$. After this first ramp, $B_\mathrm{z}$ is reduced to $-B_\mathrm{z,c} \approx -130$\,mG~$\ll-B_\mathrm{z,ramp}$ in 50\,ms. Then, the sample is evaporatively cooled to Bose condensation by reducing the dipole trapping beam intensity. To image the spin state, after switching off the trapping potential, a vertical magnetic field gradient of 8\,G/cm is applied to separate the three states in a Stern-Gerlach-like scheme. A simultaneous absorption image of the three states is made after a time-of-flight of 18\ ms. Bose condensing the sample before imaging favors the spatial spin resolution of the Stern-Gerlach imaging, given the relatively small amount of the applied magnetic field gradient and ballistic expansion time.

To reduce the sensitivity to magnetic field inhomogeneities \red{using an atomic sample with reduced spatial extension}, in Scheme B (blue line) the order between  LZ ramp and evaporation is reversed.  The atomic sample is first cooled below the critical temperature for condensation at $B_\mathrm{z,c} \approx$ 130 mG, and then the field is decreased to $B_\mathrm{z,ramp}$ to apply the LZ ramp from  $B_\mathrm{z,ramp}$ to $-B_\mathrm{z,ramp}$. 
By doing so, the spatial extent of the condensed atomic sample is considerably smaller during the LZ ramp, suppressing the contributions from the magnetic field's spatial inhomogeneities.

The ramp duration $\Delta t^*$ should be much longer than the coils time constant $\tau=L/R\simeq$ 0.5 ms, where $L$ and $R$ are the inductance and the resistance of the coils, respectively, and much shorter than the sample lifetime in the trap ($\tau_\mathrm{BEC}\simeq$ 1 s, $\tau_\mathrm{th}\gg$ 1 s). The value $B_\mathrm{z,ramp}$ was chosen depending on the ramp duration, from a maximum of 1.5\,mG to a minimum of 750\,$\mu$G, with  $|\dot{B_\mathrm{z}}|$ ranging from a maximum of 3\,G/s to a minimum of  3.75\,mG/s.

The values $B_\mathrm{x,0}$ and $B_\mathrm{y,0}$ that best compensate for the residual transverse field are the ones giving the maximal transfer to the $m_F = +1$ state. They are found through an iterative procedure, as sketched in \aref{fig:fig4}a. The first iteration is a scan in $B_\mathrm{x,coils}$ setting $|\dot{B}_\mathrm{z}|=300$\,mG/s and $B_\mathrm{y,coils} = 0$. We obtain the value $B_\mathrm{x,0}$ for which we observe maximum transfer in $m_F=+1$. Then, we set $B_\mathrm{x,coils}= B_\mathrm{x,0}$ and we perform a scan in $B_\mathrm{y,coils}$ obtaining $B_\mathrm{y,0}$ for which we have the maximal transfer of the population in the state $m_F = +1$. We repeat this procedure by slowing down the ramp in $B_\mathrm{z}$ at each iteration. 
In this way, the $P_{+1}$ distribution shrinks, increasing our sensitivity to determine the field value that best compensates for the residual one. The first iteration starts finding the value $B_\mathrm{x,0}$ with a scan in $B_\mathrm{x,coils}$ at fixed $B_\mathrm{y,coils} = 0$, with a given ramp speed $|\dot{B}_\mathrm{z}|=300$\,mG/s. Then, at fixed $B_\mathrm{x,coils}= B_\mathrm{x,0}$ we perform a scan in $B_\mathrm{y,coils}$ obtaining $B_\mathrm{y,0}$.

\begin{figure}[b!]
 \centering
  \includegraphics[width = \columnwidth]{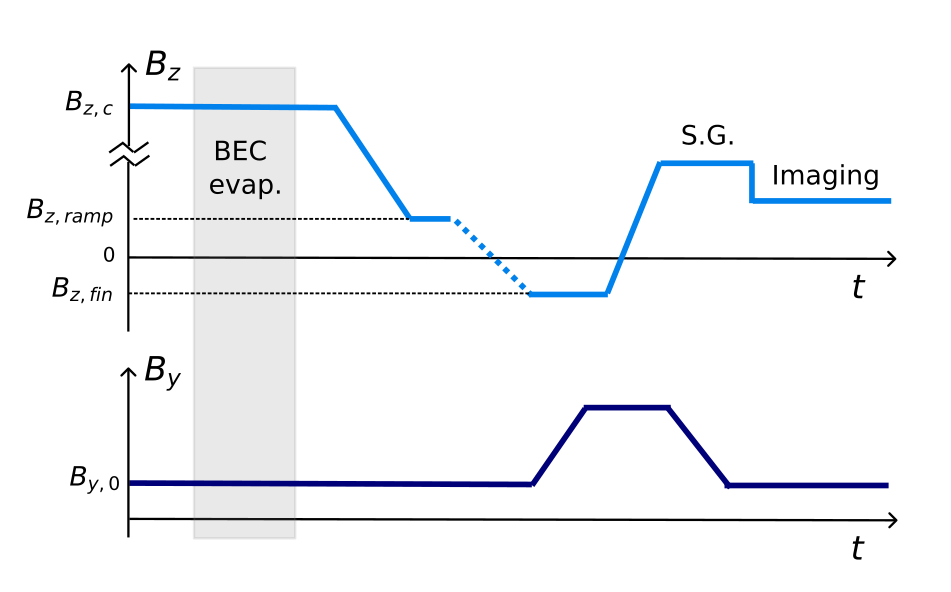}
  \caption{Longitudinal field protocol. Representation of the field in $z$ and $y$ directions. For the field in the longitudinal direction, the protocol is analogous to Scheme B with the only difference that the speed of the ramp is fixed to be $\dot{B}_\mathrm{z}=15$mG/s while the final value of the ramp $B_\mathrm{z,fin}$ is varied. The field in the $y$-direction is fixed to the optimal value $B_\mathrm{y,0}$ previously determined, it is adiabatically ramped to a finite value and then back o $B_\mathrm{y,0}$ for the imaging procedure. The field along $x$ is not shown in the figure, as it is kept at a constant value $B_\mathrm{x,0}$ for the whole duration of the experiment.}
   \label{fig:fig5}
\end{figure}

Figure~\ref{fig:fig4}b presents the three hyperfine relative populations, $n_{-1}$, $n_0$, $n_{+1}$, for three subsequent scans of the magnetic field $B_\mathrm{x,coils}$, with decreasing value of $|\dot{B}_\mathrm{z}|$ and $B_\mathrm{y,coils}$ set to the value determined in the previous iteration step. The experimental data were fitted to Eq.\,(\ref{eq:P1}-\ref{eq:Pm1}) from which we extract the center $B_\mathrm{x,0}$, the maximum transfer $P_{1,\mathrm{max}}$, and the width $\sigma_B$ of the transfer peak. At first, the center of the observed structures corresponds to $B_\mathrm{x}=0$ and this allows us to determine $B_\mathrm{x,coils}$ to compensate any residual field along $x$. 
The residual field in $y$ depends on the maximum 
in the transferred population $P_{1,\mathrm{max}}$ as
\begin{equation}
    B_y= \sigma_B \sqrt{-2\ln (P_{1,\mathrm{max}})} \, ,
\label{residualB}
\end{equation}
which is minimized after each $B_\mathrm{x}$ scan (scan in $B_\mathrm{y,coils}$ are not shown in figure). As expected, slower ramps lead to smaller $\sigma_B$ (as highlighted by the white windows, which always mark a region of 200\,$\mu$G), and consequently to increase the precision at which $B_\mathrm{x}=0$ (and $B_\mathrm{y}$) is determined.

\subsubsection{Longitudinal field}
\label{longitudinal field}

\begin{figure}[t!]
 \centering
  \includegraphics[width = \columnwidth]{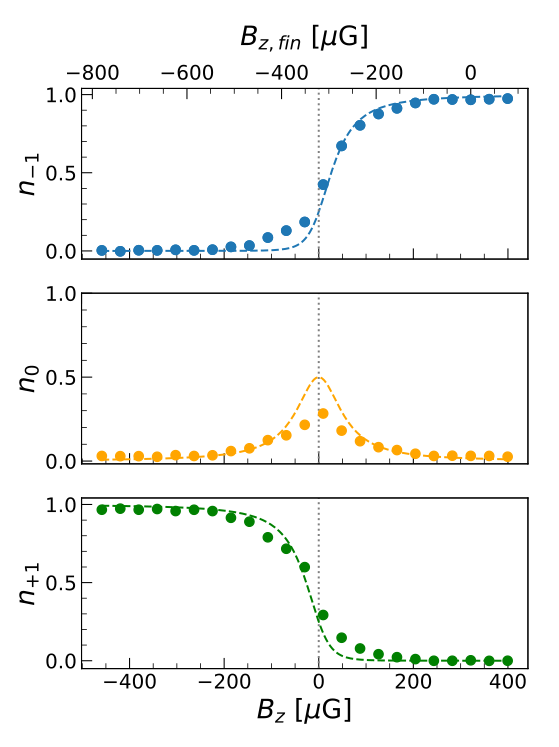}
  \caption{Example of the longitudinal field characterization. Each panel shows the relative population for the three states as a function of $B_\mathrm{z,fin}$. The dots are experimental data obtained \red{by}
  averaging different experimental realizations, with standard deviation smaller than the marker size, while dotted line is \red{the time dependent model}. The value $B_\mathrm{z,0}$ compensating for the residual \red{$B_z$ field is indicated by the dotted grey line.}}
   \label{fig:fig6}
\end{figure}

The protocol presented in the previous section allows for minimizing the transverse magnetic field. 
The best-obtained values are then used as transverse fields setting $B_\mathrm{x,coils}=B_\mathrm{x,0}$ and $B_\mathrm{y,coils}=B_\mathrm{y,0}$ for the characterization of the residual longitudinal field $B_\mathrm{z}$.

Here, we introduce the procedure to characterize the residual longitudinal field $B_\mathrm{z}$. The gas is first evaporated to obtain a BEC at $B_\mathrm{z,c}= 130$ mG, then $B_\mathrm{z,coils}$ is ramped from a positive value $B_\mathrm{z,ramp}$  down to a variable one $B_\mathrm{z,fin}$ with constant ramp $\dot{B}_\mathrm{z}=\red{-0.39}$\,G/s. As the Stern-Gerlach imaging is implemented at positive $B_\mathrm{z}$, we constrain the diabatic spin dynamics to the decreasing ramp on $B_\mathrm{z,coils}$ by raising the transverse field to a finite value after the end of the $B_\mathrm{z,coils}$ ramp, see Fig.\,\ref{fig:fig5}. If $B_\mathrm{z,fin}>0$, the LZ transfer does not take place since the ramp is interrupted before the zero crossing on $B_\mathrm{z,coils}$ hence leaving the atoms in $m_F=-1$. If $B_\mathrm{z,coils}<0$, on the other hand, the transfer to $m_F=+1$ takes place.
An example of such a scan is given in Fig.\,\ref{fig:fig6} where the relative populations in the three states are represented as a function of the value $B_\mathrm{z,fin}$. We verified that using a slower ramp does not affect the experimental result. \red{The line in the plot is a fit based on the time dependent model reported in Ref.\cite{Band2019}.  The extracted central position allows to determine $B_{z,0}$. Note that the typical oscillations of the LZ process are not visible here due to the large population transfer, which suppresses interference process between different states.}

\section{RESULTS}
\label{SectionIV}

\begin{figure}[b!]
 \centering
  \includegraphics[width = 0.9\columnwidth]{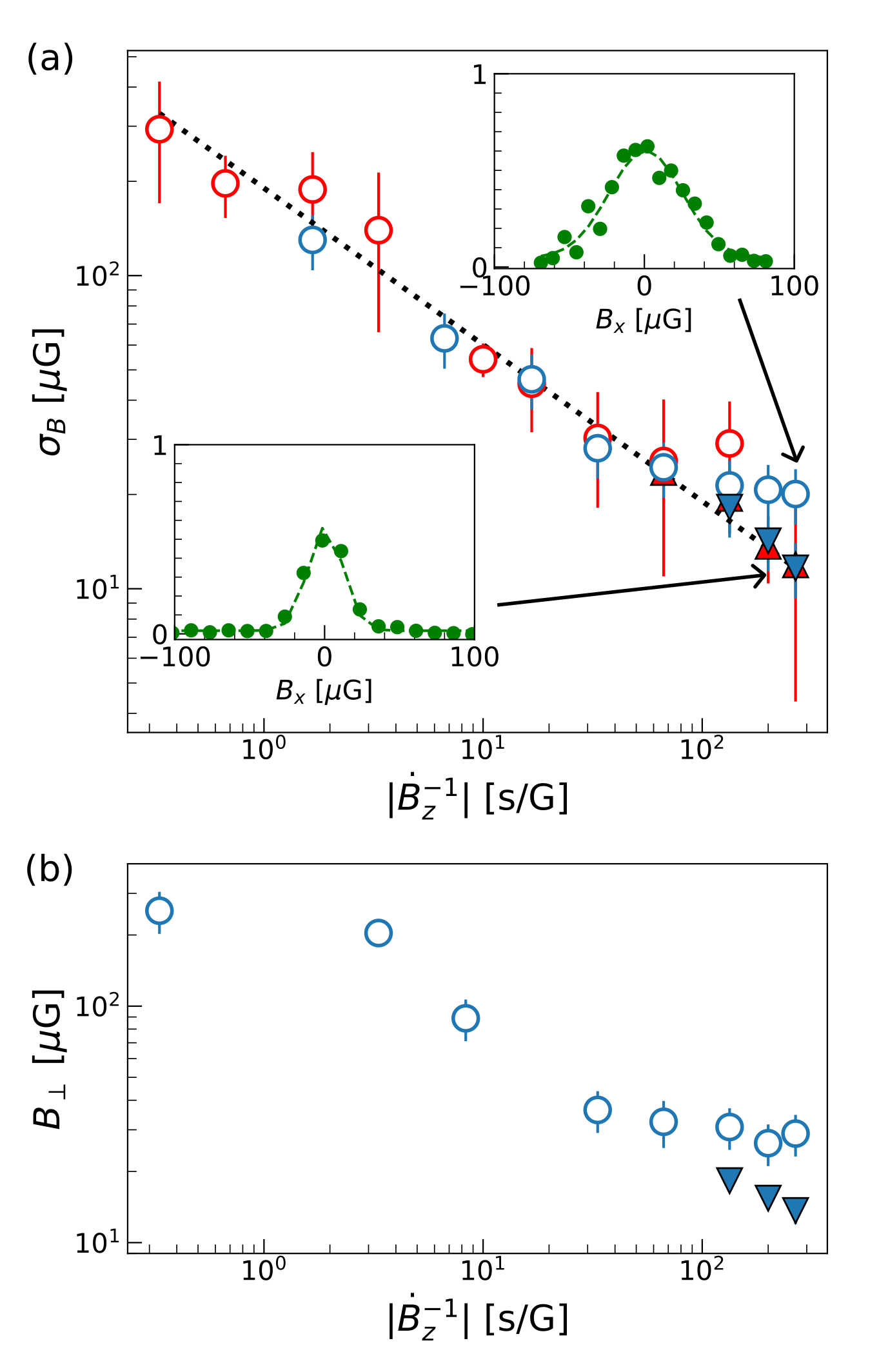}
  \caption{ a) Width $\sigma_B$ of LZ transfer probability as a function of the ramp speed. The dashed black line is the theoretical prediction from LZ theory; the symbols are the experimental results using scheme A (red symbols) or scheme B (blue symbols). Empty (filled) symbols are obtained without (with) compensation for the residual gradient, respectively. The error bars have been estimated considering the fluctuations of different datasets. The two insets are examples of the population in the state $m_F = +1$ obtained at a ramp velocity of 3.75~mG/s with (left inset) and without (right inset) the compensation of the gradient. b) Transverse residual magnetic field $B_\perp$ extracted from the population fits as a function of the inverse of the ramp velocities. Note that the error bars for the data obtained compensating the magnetic field inhomogeneities are too small to be seen in the plot.}
   \label{fig:fig7}
\end{figure}

From the fits of the experimental data, as the examples shown in Fig.\,\ref{fig:fig4}, it is possible to extract both the width of the transfer peak $\sigma_B$ and, from the contrast of the LZ population transfer, the residual transverse magnetic field $B_\perp$. 
Figure~\ref{fig:fig7}a shows the measured width $\sigma_B$ of the transfer peak as a function of $|\dot{B}_\mathrm{z}|$ for both protocols (red symbols for A and blue for B). Each point was obtained by averaging the widths from different experimental runs with the same ramp speed. The experimental values of the width are consistent with $\sigma_B$ predicted from the LZ theory (dashed line in the figure).

For instance, the flattening behavior observed in Fig.\ref{fig:fig7} around $\sigma_B=20\,\mu$G, and $B_\perp=25\,\mu$G, both for thermal and condensed samples, suggests the possible role of magnetic field inhomogeneities, in particular along the long axis of the condensate.  
We add a magnetic field gradient along the axial direction of the sample to further reduce $\sigma_B$ and $B_\perp$, as shown by the filled dots in \aref{fig:fig7}a and \aref{fig:fig7}b.

From the maximum transmission probability $P_\mathrm{1,max}$ obtained from the different scans, we can calculate $B_\perp$ by using  Eq.\,(\ref{residualB}) both for measurements without (empty symbols) and with (filled triangles) the gradient compensation, as shown in \aref{fig:fig7}b.
The smallest width observed allows us to determine the minimal residual transverse field at $(14\pm 2)\,\mu$G, a value that could be limited by the noise of the current supplies driving the compensation coils. 
Also, it is worth mentioning that the condition for the residual field compensation has generally been stable for several weeks. Still, we did observe jumps in the compensation field at the level of hundreds of $\mu$G (a few events over six months of measurements), which we could not clearly attribute to technical circumstances.

The longitudinal field was minimized by following the protocol explained in Sec.~\ref{longitudinal field}. $B_\mathrm{z,0}$ is chosen as the center of the \red{time-dependent model fitted to the data} shown in Fig.\,\ref{fig:fig6}, $ B_\mathrm{z,0} = (-323 \pm 10)\,\mu$G. 

By combining the minimal transverse and longitudinal field results, we estimate we can reach a minimal field modulus of $(18\pm 5)\,\mu$G, which complies with the conditions for observing a nematic phase in an elongated \na condensate.

\section{CONCLUSIONS AND OUTLOOK}
\label{SectionV}

In this paper, we present a technique to characterize and compensate magnetic fields at the level of 10 $\mu$G in experiments with ultracold atomic gases. The method is based on monitoring the Zeeman populations in diabatic atomic spin rotation. The simulations based on the LZ dynamics for a three-level system reproduced the experimental data well. These results pave the way to studying the unexplored scenario of condensation in zero magnetic fields in extended spinor gases, when spin interactions may become dominant over all other contributions to the Hamiltonian. 
For instance, the ground state of an F=1 spinor condensate, characterized by antiferromagnetic interactions, develops an order parameter with a nematic character, as observed in the single spatial mode approximation \cite{Hamley2012, Zibold2016}, but its superfluid properties were not explored so far.\\

\section*{Acknowledgements}
We acknowledge funding from Provincia Autonoma di Trento, from the European Union’s Horizon 2020 research and innovation Programme through the STAQS project of QuantERA II (Grant Agreements No. 101017733)
and from the European Union - Next Generation EU through PNRR MUR project PE0000023-NQSTI.
The work was also supported by the  project DYNAMITE QUANTERA2-00056, funded by the Ministry of University and Research through the ERANET COFUND QuantERA II – 2021 call and co-funded by the European Union (Grant Agreements No. 101017733).  This work was supported by Q@TN, the joint lab between the University of Trento, FBK - Fondazione Bruno Kessler, INFN - National Institute for Nuclear Physics and CNR - National Research Council.


%

\end{document}